\documentclass{article}          
\usepackage{epsfig}
\usepackage[dvips]{color}
\usepackage{geometry}
\begin{document}

\newcommand{\eins}{\mbox{$1 \hspace{-1.0mm}  {\bf l}$}}
\newcommand{\D}[3]{D_{#1:#2}(#3)}
\newcommand{\Done}[2]{D_{#1}(#2)}
\newcommand{\s}[2]{\sigma_{#1}^{(#2)}}
\newcommand{\im}{{\rm i}}
\newcommand{\Clus}{{\cal C}}
\newcommand{\half}{{\frac{1}{2}}}
\newcommand{\ket}[1]{|#1\rangle}
\newcommand{\bra}[1]{\langle#1|}
\newcommand{\braket}[2]{\langle#1|#2\rangle}
\newcommand{\vek}[1]{{\boldsymbol {#1}}}
\newcommand{\iden}{\eins\,}

\newcommand{\ncd}{\newcommand}
\ncd{\x}{$\bullet\,\,\,\,$}
\ncd{\oo}{$\mbox{}\,\,\,\,\,\,\,$}
\ncd{\nil}{$\bigcirc$}
\ncd{\pu}{$\bullet$}
\ncd{\ua}{$\uparrow$}
\ncd{\ra}{$\rightarrow$}
\ncd{\ds}{\displaystyle}
\ncd{\dummy}{\mbox{\tiny{\textcolor[cmyk]{0.5,0,0,0}{+}}}}
\ncd{\CNOT}{\mbox{CNOT}}
\ncd{\QC}{$\mbox{QC}_{\cal{C}}\,\,$}
\ncd{\QCns}{$\mbox{QC}_{\cal{C}}$}
\ncd{\fc}{\mbox{fc}}
\ncd{\bc}{\mbox{bc}}
\ncd{\notexists}{\exists \!\!\hspace*{-0.28mm}|\,\,\hspace*{0.28mm}}

\title{The one-way quantum computer -- a non-network model of quantum
  computation }

\author{Robert Raussendorf, \thanks{email:
    raussen@theorie.physik.uni-muenchen.de}\ \mbox{ } 
    Daniel E. Browne \thanks{email:
    browne@theorie.physik.uni-muenchen.de}\ \mbox{ } and 
    Hans J. Briegel \thanks{email:
    briegel@theorie.physik.uni-muenchen.de} \\
    Ludwig-Maximilians-Universit{\"a}t M{\"u}nchen}
  
\maketitle

\begin{abstract}
   A one-way quantum computer (\QCns) works by only performing a sequence of
   one-qubit measurements on a particular entangled multi-qubit state,
   the cluster state. No non-local operations are
   required in the 
   process of computation. Any quantum logic network can be simulated
   on the \QCns. On the other hand, the network model of quantum
   computation cannot explain all ways of processing quantum information
   possible with the \QCns. In this paper, two examples of
   the non-network 
   character of the \QC are given. First, circuits in the Clifford
   group can be performed in a single time 
   step. Second, the \QCns-realisation of a particular circuit
   --the bit-reversal gate-- has no network interpretation.
\end{abstract}

\section{Introduction}

Recently, we introduced the concept of the ``one-way quantum
computer'' in \cite{QCmeas}. In this scheme, the whole quantum
computation consists only of a sequence of one-qubit projective
measurements on a given entangled state, a so called cluster state
\cite{BR}. We called this scheme the ``one-way
quantum computer'' since the entanglement in a cluster state is
destroyed by the one-qubit measurements and therefore the cluster state
can be used only once. In this way, the cluster state forms a resource 
for quantum computation. The set of measurements form the
program. To stress the importance of the cluster state 
for the scheme, we will use in the following the abbreviation \QC for
``one-way quantum computer''. 

As we have shown in \cite{QCmeas}, any quantum logic network
\cite{QLNW} can be
simulated on the \QCns. On the other hand, the quantum logic network
model cannot 
explain all ways of quantum information processing that are possible
with the \QCns. Circuits that realise transformations
in the Clifford group  
--which is generated by all the  CNOT-gates,
Hadamard-gates and $\pi/2$ phase shifts-- can be performed
by a \QC in a single step, i.e. all the measurements to implement
such a circuit can be carried out at the same time. The best networks
that have been found  have 
a logical depth logarithmic in the number of qubits \cite{M&N}.   In a
simulation 
of a quantum logic network by a one-way quantum computer, the temporal
ordering of the gates of the network is transformed into a spatial
pattern of the measurement directions on the resource cluster
state. For the temporal ordering of the measurements there seems to be no
counterpart in the network model. Further, not every measurement
pattern that implements a larger circuit can be decomposed into smaller
units, as can be seen from the example of the bit-reversal gate given
in Section~\ref{brg}.
 
The purpose of this paper is to illustrate the 
non-network character of the \QC using two examples -- first,
the temporal complexity of circuits in the Clifford group, and second,
the bit-reversal gate on the \QC which has no network
interpretation.

\section{Summary of the one-way quantum computer}
\label{summary}

In this section, we give an outline of the universality proof
\cite{QCmeas} for the \QC. For the one-way quantum computer, the
entire resource for the quantum 
computation is provided  initially in the form of a specific entangled
state --the cluster state \cite{BR}-- 
of a large number of qubits. Information is then written onto the 
cluster, processed, and read out from the cluster by one-particle 
measurements only. The entangled state of the cluster thereby serves as a 
universal ``substrate'' for any quantum computation. Cluster states can be created 
efficiently in any system with a quantum Ising-type interaction (at very low 
temperatures) between two-state particles in a lattice
configuration. More specifically, to create a cluster state
$|\phi\rangle_{\cal{C}}$, the qubits on a cluster ${\cal{C}}$ 
are at first all prepared individually in a state $|+\rangle = 1/\sqrt{2}
(|0\rangle + |1\rangle)$ and then brought into a cluster state by
switching on the Ising-type interaction $H_{\mbox{\tiny{int}}}$ for an
appropriately chosen finite time span $T$. The time evolution operator
generated by this Hamiltonian which takes the initial
product state to the cluster state is denoted by $S$.

The quantum state $|\phi\rangle_{\cal{C}}$, the cluster state of a 
cluster $\cal C$ of neighbouring qubits, provides in advance all
entanglement that is involved in the subsequent quantum  
computation. It has been shown \cite{BR} 
that the cluster state $|\phi\rangle_{\cal C}$ is characterised by a
set of eigenvalue  equations
\begin{equation}
    \sigma_x^{(a)} \bigotimes_{a'\in ngbh(a)}\sigma_z^{(a')} 
    |\phi\rangle_{\cal C} = \pm |\phi\rangle_{\cal C}, 
\label{EVeqn}
\end{equation} 
where $ngbh(a)$ specifies the sites of all qubits 
that interact with the qubit at site $a\in {\cal C}$. The eigenvalues
are determined by 
the distribution of the qubits on the lattice. The equations (\ref{EVeqn}) are
central for the proposed computation scheme. It is important to realise
here that information processing is possible even though the result of
every measurement in any direction of the Bloch sphere is completely
random. The reason for the randomness of the measurement results is
that the reduced density operator for 
each qubit in the cluster state is $\frac{1}{2}{\bf{1}}$. While the
individual measurement results are irrelevant for the computation, the
strict correlations between measurement results inferred from 
(\ref{EVeqn}) are what makes the processing of quantum information
possible on the \QCns.

For clarity, let us emphasise that in the scheme of the \QC  we
distinguish between cluster qubits on 
${\cal{C}}$  which are measured in the process of computation, and the
logical qubits. The  
logical qubits constitute the quantum information being processed while
the cluster qubits in the initial cluster state form an entanglement resource.
Measurements of their individual one-qubit state drive the
computation.

\begin{figure}[tph]
\begin{center}
 \epsfig{file=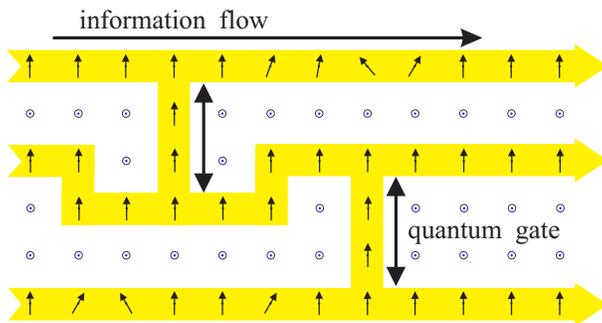,width=8cm}
 \parbox{0.7\textwidth}{\caption{\label{FIGetching}Quantum computation
 by measuring 
 two-state particles on a 
  lattice. Before the measurements the qubits are in the
  cluster state $|\phi\rangle_{\cal{C}}$ of (\ref{EVeqn}). 
  Circles $\odot$ symbolise measurements of $\sigma_z$, vertical arrows
  are measurements of $\sigma_x$, while tilted arrows refer to
  measurements in the x-y-plane.}}

\end{center}
\end{figure}

To process quantum information with this cluster, it suffices to
measure its particles in a certain order and in a certain basis, as
depicted in Fig.~\ref{FIGetching}. Quantum 
information is thereby propagated through the cluster and
processed. Measurements of $\sigma_z$-observables effectively remove the
respective lattice qubit from the cluster. Measurements in the
$\sigma_x$-eigenbasis are
used for ``wires'', i.e. to propagate logical quantum bits through the
cluster, and for the CNOT-gate between two logical qubits. Observables
of the form $\cos(\varphi)\,\sigma_x \pm \sin(\varphi)\, \sigma_y$ are
measured to realise arbitrary rotations of logical qubits. 
For cluster qubits to implement rotations, the basis in  which each of
them is  measured depends on the results of preceding 
measurements. This introduces a temporal order in which the
measurements have to be performed. The processing is finished once all qubits 
except a last one on each wire have been measured. The remaining
unmeasured qubits form the quantum register which is now ready to be
read out. At this point, the
results of previous measurements  
determine in which basis these ``output'' qubits need to be measured for the 
final readout, or if the readout measurements are in the $\sigma_x$-,
$\sigma_y$- 
or $\sigma_z$-eigenbasis, how the readout measurements have to be
interpreted. Without loss of generality, we assume in this paper that
the readout measurements are performed in the $\sigma_z$-eigenbasis. 

Here, we review two points of the
universality proof for the \QCns: 
the realisation of the arbitrary one-qubit rotation as a member of the
universal set of gates, and the effect of the randomness of
the individual measurement results and how to account for them. For
the realisation of a CNOT-gate see Fig.~\ref{Gates} and \cite{QCmeas}.
   
An arbitrary rotation $U_R \in SU(2)$ can be achieved in a
chain of 5 qubits. Consider a
rotation in its Euler representation 
\begin{equation}
    \label{Euler}
    U_R(\xi,\eta,\zeta) = U_x(\zeta)U_z(\eta) U_x(\xi),
\end{equation}
where the rotations about the $x$- and $z$-axis are 
$
        U_x(\alpha) =
        \displaystyle{\mbox{exp}\left(-i\alpha\frac{\sigma_x}{2}\right)} 
$ and
$  
        U_z(\alpha) = \displaystyle{\mbox{exp}\left(-i\alpha 
        \frac{\sigma_z}{2}\right)} 
$.
Initially, the
first qubit is in some state
$|\psi_{\mbox{\footnotesize{in}}}\rangle$, which is to be rotated, and
the other qubits are in 
$|+\rangle$. After the 5 qubits are entangled by the time
evolution operator $S$ generated by the Ising-type Hamiltonian, the
state $|\psi_{\mbox{\footnotesize{in}}}\rangle$ can be rotated by measuring qubits 1 to
4. At the same time, the state is also transfered to site 5. The qubits $1
\dots 4$ are measured in appropriately  chosen bases, {\em{viz.}}
\begin{equation}
    \label{Measbas}
    {\cal{B}}_j(\varphi_{j}) = \left\{
        \frac{|0\rangle_j+e^{i \varphi_{j}} |1\rangle_j}{\sqrt{2}} ,\,
        \frac{|0\rangle_j-e^{i \varphi_{j}} |1\rangle_j}{\sqrt{2}}
    \right\}
\end{equation} 
whereby the measurement outcomes  $s_{j} \in \{ 0,1 \}$ for
$j=1\dots 4$ are obtained. Here, $s_{j}=0$ means that qubit $j$ is projected 
into the first state of
${\cal{B}}_j(\varphi_{j})$. In
 (\ref{Measbas}) the basis states of all possible measurement bases
lie on the equator of the Bloch sphere, i.e. on the intersection of
the Bloch sphere with the $x-y$-plane. Therefore, the measurement
basis for qubit $j$ can be specified by a single parameter, the
measurement angle $\varphi_{j}$. The
measurement direction of qubit $j$ is the vector on the Bloch sphere
which corresponds 
to the first state in the measurement basis
${\cal{B}}_j(\varphi_{j})$. Thus, the
measurement angle $\varphi_{j}$ is equal to the
angle between the measurement direction at qubit $j$ and the positive
$x$-axis. For all of
the gates constructed so far, the cluster qubits are either 
--if they are not required for the realisation of the
circuit-- measured in $\sigma_z$, or --if they are required-- measured
in some measurement direction in the $x-y$-plane. 
In summary,
the procedure to implement an arbitrary rotation $U_R(\xi,\eta,\zeta)$, 
specified by its Euler angles $\xi,\eta,\zeta$, is this:
1. measure qubit 1 in ${\cal{B}}_1(0)$; 2. measure qubit 2 in
 ${\cal{B}}_2\left( (-1)^{s_1+1} \xi \right)$; 3. measure qubit 3 in
 ${\cal{B}}_3\left( (-1)^{s_2} \eta \right)$; 4. measure qubit 4 in 
 ${\cal{B}}_4\left( (-1)^{s_1+s_3} \zeta \right)$. In this way the
rotation $U_R^\prime$ is realised:
\begin{equation}
    \label{Rotprime}
    U_R^\prime(\xi,\eta,\zeta) =  U_\Sigma U_R(\xi,\eta,\zeta).
\end{equation}
The random byproduct operator 
\begin{equation}
    \label{Byprod1}
    U_\Sigma= \sigma_x^{s_2+s_4}\sigma_z^{s_1+s_3}
\end{equation} can
be corrected for at the end of the computation, as explained next.  

The randomness of the measurement results does not
jeopardise the function of the  circuit. Depending on
the measurement results, extra rotations $\sigma_x$ and $\sigma_z$ act on
the output qubits of every implemented gate, as in (\ref{Rotprime}), for
example. By use of the propagation
relations
\begin{eqnarray}
    \label{Rotprop}
    U_R(\xi,\eta,\zeta) \, \sigma_z^{s} \sigma_x^{s'} &=&   
    \sigma_z^{s} \sigma_x^{s'} \,
    U_R((-1)^{s}\xi,(-1)^{s'}\eta,(-1)^{s}\zeta),\\ 
    \label{CNTprop}
    \mbox{CNOT} (c,t) \, {\sigma_z^{(t)}}^{s_t} {\sigma_z^{(c)}}^{s_c}
    {\sigma_x^{(t)}}^{s_t'} {\sigma_x^{(c)}}^{s_c'} 
    &=&  {\sigma_z^{(t)}}^{s_t} {\sigma_z^{(c)}}^{s_c+s_t}
    {\sigma_x^{(t)}}^{s_c'+s_t'} {\sigma_x^{(c)}}^{s_c'} \, 
    \mbox{CNOT}(c,t) ,
\end{eqnarray}  
these extra rotations can be pulled
through the network to act upon the output state. There they can be
accounted for by properly interpreting the $\sigma_z$-readout
measurement results.

To summarise, any quantum logic network can be
simulated on a one-way quantum computer. A set of universal gates can
be realised by one-qubit measurements and the gates can be combined to
circuits. Due to the randomness of the results of the individual
measurements, unwanted byproduct operators are introduced. These
byproduct operators can be accounted for by 
adapting measurement directions throughout the process. In this way, a
subset of qubits on the cluster ${\cal{C}}$ is prepared as the output
register. The quantum state on this subset of qubits equals
that of the quantum register of the simulated network up to the action
of an 
accumulated byproduct operator. The byproduct operator determines how the
measurements on the output register are to be interpreted.

\section{Non-network character of the \QC}

\subsection{Logical depth $D=1$ for circuits in the Clifford group}
\label{Clifford}

The Clifford group of gates is generated by the CNOT-gates, the
Hadamard-gates and the $\pi/2$-phase shifts. 
In this section it is proved that the logical depth of
circuits belonging to the Clifford group is $D=1$ on
the \QCns, irrespective of the number of
logical qubits $n$. For a subgroup of the Clifford
group, the group generated by the CNOT- and Hadamard gates we can
compare the result to the best known upper bound for quantum logic 
networks where the logical depth scales like $O(\log n)$ \cite{M&N}. 

\begin{figure}
    \begin{center}
        \begin{tabular}{ccc}
            \parbox{5.7cm}{\epsfig{file=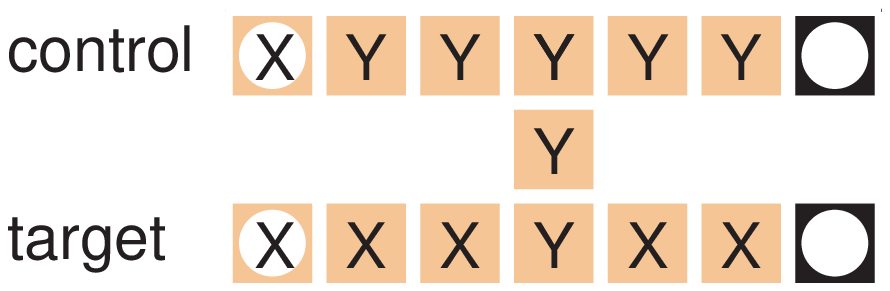,width=5.7cm}} &
            \parbox{3.2cm}{\epsfig{file=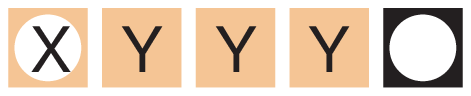,width=3.2cm}} &
            \parbox{3.2cm}{\epsfig{file=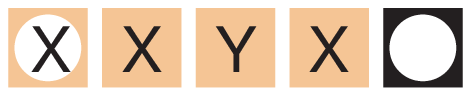,width=3.2cm}} \\
            CNOT-gate & Hadamard-gate &
            $\pi/2$-phase gate 
           
        \end{tabular}
        \parbox{0.8\textwidth}{\caption{\label{Gates}{Realisation of 
              the required gates on the \QCns.  CNOT-gate between
              neighbouring qubits, the Hadamard gate and the $\pi/2$
              phase gate. }}}
    \end{center}
\end{figure}
The Hadamard- and the $\pi/2$-phase gate are, compared to general
$SU(2)$-rotations, special with regard to the measurements which are
performed on 
the cluster to implement them. The Euler angles (\ref{Euler}) that implement a
Hadamard- and a $\pi/2$-phase gate are, depending on the byproduct
operator on the input side, given by $\xi=\pm \pi/2,\,\, \eta=\pm
\pi/2, \,\,
\zeta=\pm \pi/2$ and $\xi=0, \,\,\eta=\pm \pi/2,\,\, \zeta=0$,
respectively. See Fig.~\ref{Gates}. A
measurement angle of 0 corresponds to a measurement of the observable
$\sigma_x$ and both  measurement angles $\pi/2$ and $-\pi/2$
correspond to a measurement of $\sigma_y$. A change of the measurement
angle from $\pi/2$ to $-\pi/2$  has only the effect
of interchanging 
the two states of the measurement basis in (\ref{Measbas}), but it does not change the basis
itself. 

In general, cluster qubits which are measured in the eigenbasis of
$\sigma_x$, $\sigma_y$ or $\sigma_z$ can be measured at the same time. The
adjustment of their measurement basis does not require classical
information gained by measurements on other cluster qubits. As
described in \cite{QCmeas}, $\sigma_z$-measurements are used to
eliminate those cluster qubits which are not essential for the
circuit. As shown in Fig.~\ref{Gates}, the cluster qubits essential
for the implementation of the CNOT-gate between neighbouring qubits,
the Hadamard- and the 
$\pi/2$-phase gate are all measured in the $\sigma_x$- or
$\sigma_y$-eigenbasis. The general CNOT can be constructed from
CNOT-gates between neighbouring qubits.  This can be done in the standard
manner using  the swap-gate composed of three
CNOT-gates. Hence also the general CNOT-gate requires only $\sigma_x$-
and $\sigma_y$-measurements. Therefore, all circuits in the Clifford
group can be  realised in a single
step of parallel measurements.

\subsection{The bit-reversal gate}
\label{brg}

\begin{figure}
\begin{center}
\epsfig{file=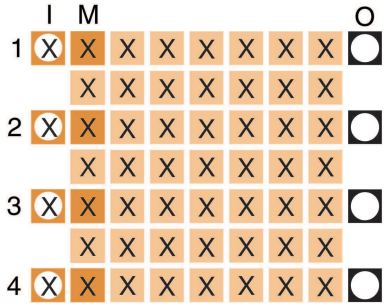,width=5cm}
\parbox{0.7\textwidth}{\caption{This measurement pattern implements a
    bit-reversal gate. Each square represents a lattice qubit. The
    squares in the extreme left column marked with white circles
    denote the input qubits, they on the right-most column denote the
    output qubits. Note that blank squares can represent either
    $\sigma_z$ measurements or  empty lattice sites.}\label{bitrev-fig}}
\end{center}
\end{figure}

\noindent
The measurement pattern shown in Fig.~\ref{bitrev-fig} realises a
bit-reversal gate. In this example, it acts on four logical qubits,
reversing the bit 
order, $\ket{x_3 x_2  x_1 x_0}\to \ket{x_0 x_1 x_2 x_3}$. 
Reconsidering the circuit of CNOT-gates and one-qubit rotations
depicted in Fig.~\ref{FIGetching}, the network structure --displayed
in gray underlay-- is clearly reflected in the measurement
pattern. One finds the wires for logical 
qubits ``isolated'' from each other by areas of qubits 
measured in $\sigma_z$ and ``bridges'' between the wires which realise
two-qubit gates.

For the bit-reversal gate in Fig.~\ref{bitrev-fig}  the situation is
different. To implement this gate 
on $n$ logical qubits, the cluster qubits on a square block of size
$2n-1\, \times 2n-1$ all have to be measured in the
$\sigma_x$-eigenbasis. The circuit which is realised via this
measurement pattern cannot be explained by decomposing it into smaller
parts. There is no network interpretation --as there is for
Fig.~\ref{FIGetching}-- for the measurement pattern  displayed in
Fig.~\ref{bitrev-fig}. 

The bit reversal gate can be understood in a similar way to
teleportation. Here we give an explanation for the bit-reversal gate
on four qubits, but the argument can be straightforwardly generalised to an
arbitrary number $n$ of logical qubits. In Fig.~\ref{bitrev-fig} we
have the set of input qubits $I=\{i_1,\dots, i_4\}$, their
neighbouring qubits $M=\{m_1,\dots,m_4\}$ and the set of output qubits
$O=\{o_1,\dots,o_4\}$. All the qubits required for the gate form the 
set $Q$. Further, we define the ``body'' G of the gate as the set $G=Q
\backslash (I\cup M \cup O)$. As in \cite{QCmeas}, the standard
setting to discuss a gate is the following: 1.) Prepare the input
qubits $I$ in an input state $|\psi_{\mbox{\footnotesize{in}}} \rangle$
and the remaining qubits each in $|+\rangle$. 2.) Entangle the qubits
in  $Q$ by $S$, the unitary transformation generated by the Ising-interaction
$H_{\mbox{\tiny{int}}}$. 3.) Measure the qubits in $I \cup M \cup G$,
i.e. all but the output qubits, in the $\sigma_x$-eigenbasis. 
Due to linearity we can confine ourselves to input states of product form
$|\psi_{\mbox{\footnotesize{in}}} \rangle= |\alpha\rangle_{i_1}\otimes
 |\beta\rangle_{i_2}\otimes  |\gamma\rangle_{i_3}\otimes
 |\delta\rangle_{i_4}.$ The output state we get after step 3
 in the above protocol is
 $|\psi_{\mbox{\footnotesize{out}}} \rangle= U_\Sigma\,\,
 |\delta\rangle_{o_1} \otimes |\gamma \rangle_{o_2} \otimes
 |\beta\rangle_{o_3} \otimes  |\alpha\rangle_{o_4}$. There, the
 multi-local byproduct operator $U_\Sigma$ belongs to a discrete set
 and depends only on the results of the measurements.

The entanglement operation $S$ on the qubits in $Q$ can be written as
a product $S=S_{MGO}\, S_{IM}$ where $S_{MGO}$ entangles the qubits in
$M$,$G$ and $O$ and $S_{IM}$ entangles the input qubits in $I$ with
their neighbours in $M$. $S_{MGO}$ and $S_{IM}$ commute. Further,
let $P$ denote the projection operator which describes the set of
measurements in step 2 of the above protocol. Then, the projector $P$
can be written as a product as well, $P= P_G\, P_{IM}$. There, $P_G$
denotes the projector associated with the measurements on the set $G$
of qubits, and $P_{IM}$ the projector associated with the measurements
on $I$ and $M$. Now note that $P_G$ commutes with $S_{IM}$,
because the two operators act non-trivially only  on different subsets of
qubits. Therefore, the above protocol is mathematically equivalent to
the following 
one: $1^\prime$.) Prepare the input
qubits $I$ in an input state $|\psi_{\mbox{\footnotesize{in}}} \rangle$
and the remaining qubits each in $|+\rangle$. $2^\prime$.) Entangle the qubits
in  $M$, $G$ and $O$ by $S_{MGO}$. $3^\prime$.) Measure the qubits in
$G$. $4^\prime$.) 
Entangle the qubits in $I$ and $M$ by $S_{IM}$. And $5^\prime$.) Measure the
qubits in $I$ and $M$.  

The quantum state of the qubits in $I$, $M$ and $O$ after step $3^\prime$ in
the second protocol is of the form
$|\Psi\rangle_{IMO}=|\psi_{\mbox{\footnotesize{in}}}\rangle_I \otimes
|\Phi\rangle_{MO}$. As a consequence of the eigenvalue equations
(\ref{EVeqn}) of the unmeasured state of the qubits in $M$, $G$ and
$O$, the state $|\Phi\rangle_{MO}$ obeys the following set of eight
eigenvalue equations: $\sigma_x^{(m_1)}\sigma_z^{(o_4)} \,
|\Phi\rangle_{MO} = \pm |\Phi\rangle_{MO}$,
$\sigma_z^{(m_1)}\sigma_x^{(o_4)} \, 
|\Phi\rangle_{MO} = \pm |\Phi\rangle_{MO}$, and the three remaining pairs of
equations with $(m_1,o_4)$ replaced by $(m_2,o_3)$, $(m_3,o_2)$ and
$(m_4,o_1)$. The sign factors in these equations depend on the
results measured in step $3^\prime$ of the second protocol. The eigenvalue
equations determine the state $|\Phi\rangle_{MO}$
completely. It is the product of four Bell states
$|\mbox{B}\rangle$: 
$|\Phi\rangle_{MO} =_{\mbox{\scriptsize l.u.}} |\mbox{B}\rangle_{m_1o_4}
\otimes   |\mbox{B}\rangle_{m_2o_3}
\otimes   |\mbox{B}\rangle_{m_3o_2}
\otimes   |\mbox{B}\rangle_{m_4o_1}$, where $|\mbox{B}\rangle =
1/\sqrt{2}(|+\rangle |1\rangle + |-\rangle |0\rangle)$ and ``$ =_{\mbox{\scriptsize
    l.u.}}$'' means 
equal up to possible local unitarities $\sigma_x, \sigma_z$ on the
output qubits in $O$. The entanglement operation $S_{IM}$ in step
$4^\prime$ together with the local 
measurements $P_{IM}$ in step $5^\prime$ has the effect of four Bell
measurements on the qubit pairs $(i_k,m_k), \; k=1\dots4$, in the basis
$\{1/\sqrt{2}(|0\rangle_{i_k} |\pm\rangle_{m_k} \pm |1\rangle_{i_k}
|\mp\rangle_{m_k}) \}$. 

The second
protocol --which is mathematically equivalent to the first-- is thus a
teleportation scheme. In steps $2^\prime$ and $3^\prime$ Bell pairs
between intermediate 
qubits $m_k \in M$ and the output qubits are created. The first qubit
in $M$ forms a Bell state with the last qubit in $O$, and so on: 
\begin{center}
    \vspace*{-0.3cm} 
    \setlength{\unitlength}{0.6cm}
    \setlength{\unitlength}{0.6cm}
    \begin{picture}(6,5)

        \color{white}
        \put(2.5125,3.5125){\line(1,-1){3}}
        \put(2.4875,3.4875){\line(1,-1){3}}
        \color{black}
        \linethickness{0.25mm}
        \put(2.5,3.5){\line(1,-1){3}}

        \color{white}
        \put(2.5056,2.5145){\line(3,-1){3}}
        \put(2.4944,2.4865){\line(3,-1){3}}
        \color{black}
        \put(2.5,2.5){\line(3,-1){3}}

        \color{white}
        \put(2.5056,1.4865){\line(3,1){3}}
        \put(2.4944,1.5145){\line(3,1){3}}
        \color{black}
        \put(2.5,1.5){\line(3,1){3}}

        \color{white}
        \put(2.5125,0.4875){\line(1,1){3}}
        \put(2.4875,0.5125){\line(1,1){3}}
        \color{black}
        \put(2.5,0.5){\line(1,1){3}}

        \put(1.5,0.5){\circle*{0.7}}
        \put(1.5,1.5){\circle*{0.7}}
        \put(1.5,2.5){\circle*{0.7}}
        \put(1.5,3.5){\circle*{0.7}}

        \put(2.5,0.5){\circle*{0.7}}
        \put(2.5,1.5){\circle*{0.7}}
        \put(2.5,2.5){\circle*{0.7}}
        \put(2.5,3.5){\circle*{0.7}}

        \put(5.5,0.5){\circle*{0.7}}
        \put(5.5,1.5){\circle*{0.7}}
        \put(5.5,2.5){\circle*{0.7}}
        \put(5.5,3.5){\circle*{0.7}}

        \color{white}
        \put(1.5,0.5){\circle*{0.45}}
        \put(1.5,1.5){\circle*{0.45}}
        \put(1.5,2.5){\circle*{0.45}}
        \put(1.5,3.5){\circle*{0.45}}

        \put(2.5,0.5){\circle*{0.45}}
        \put(2.5,1.5){\circle*{0.45}}
        \put(2.5,2.5){\circle*{0.45}}
        \put(2.5,3.5){\circle*{0.45}}

        \put(5.5,0.5){\circle*{0.45}}
        \put(5.5,1.5){\circle*{0.45}}
        \put(5.5,2.5){\circle*{0.45}}
        \put(5.5,3.5){\circle*{0.45}}

        \color{black}
        \put(0,3.37){${|\alpha\rangle}$}
        \put(0,2.37){${|\beta\rangle}$}
        \put(0,1.37){${|\gamma\rangle}$}
        \put(0,0.37){${|\delta\rangle}$}

        \put(1.39,4.1){$I$}
        \put(2.2,4.1){$M$}
        \put(5.39,4.1){$0$}

\end{picture}
    \vspace*{-0.2cm}
\end{center}
In
steps $4^\prime$ and $5^\prime$ a Bell measurement on each pair of
qubits $(i_k,m_k)$ is 
performed. In this way, the input state
$|\psi_{\mbox{\footnotesize{in}}}\rangle$ is teleported from the
input to the output qubits, with the order of the qubits  
reversed. As in teleportation, the output state is equivalent to the
input state only up to the action of a multi-local unitary operator
which is specified by two classical bits per teleported qubit. In the
case of the bit-reversal 
gate the role of this operator is taken by the byproduct operator $U_\Sigma$. 
  
Please note that there is no temporal order of measurements
here. {\em{All measurements can be carried out at the same time.}} The apparent
temporal order in the second
protocol was introduced only as a pedagogical
trick to explain
the bit-reversal gate in terms of teleportation.

\section{Conclusion}

In this paper, we have shown that the network picture cannot
describe all ways of quantum information processing that are possible
with a one-way 
quantum computer. First, circuits which realise unitary
transformations in the Clifford group can be performed on the \QC in a
single time 
step. This includes in particular all circuits composed of CNOT- and
Hadamard gates, for which the best known network has a 
logical depth that scales logarithmically in the number of qubits. Second,
we presented an example of a circuit on the \QCns, the bit-reversal
gate, which cannot be interpreted as a network composed of gates. 
These observations motivate a novel computational model underlying the
\QCns, which is described in \cite{model}.

\section*{Acknowledgements}

This work has been supported by the Deutsche Forschungsgemeinschaft
(DFG) within the Schwerpunktprogramm QIV and by the Deutscher
Akademischer Austauschdienst (DAAD). We would like to thank O. Forster for
helpful discussions.

\end{document}